**Giant birefringence in optical antenna arrays with widely tailorable optical anisotropy**

**Mikhail A. Kats[1], Patrice Genevet[1,2], Guillaume Aoust[1,3], Nanfang Yu[1], Romain Blanchard[1], Francesco Aieta[1,4], Zeno Gaburro[1,5], and Federico Capasso[1,*]**

[1]School of Engineering and Applied Sciences, Harvard University, Cambridge, Massachusetts 02138, USA

[2]Institute for Quantum Studies and Department of Physics, Texas A&M University, College Station, Texas 77843, USA

[3]Department of Physics, Ecole Polytechnique, Paris, France

[4]Dipartimento di Fisica e Ingegneria dei Materiali e del Territorio, Università Politecnica delle Marche, via Brecce Bianche, 60131 Ancona, Italy

[5]Dipartimento di Fisica, Università degli Studi di Trento, via Sommarive 14, 38100 Trento, Italy

Classification: Physical Sciences: Applied Physical Sciences

*Corresponding author:

Federico Capasso, capasso@seas.harvard.edu (1-617-384-7611)
Pierce 205A, 29 Oxford Street, Cambridge, MA 02138

**Abstract:**

**The manipulation of light by conventional optical components such as a lenses, prisms and wave plates involves engineering of the wavefront as it propagates through an optically-thick medium. A new class of ultra-flat optical components with high functionality can be designed by introducing abrupt phase shifts into the optical path, utilizing the resonant response of arrays of scatters with deeply-subwavelength thickness. As an application of this concept, we report a theoretical and experimental study of birefringent arrays of two-dimensional (V- and Y-shaped) optical antennas which support two orthogonal charge-oscillation modes and serve as broadband, anisotropic optical elements that can be used to locally tailor the amplitude, phase, and polarization of light. The degree of optical anisotropy can be designed by controlling the interference between the light scattered by the antenna modes; in particular, we observe a striking effect in which the anisotropy disappears as a result of destructive interference. These properties are captured by a simple, physical model in which the antenna modes are treated as independent, orthogonally-oriented harmonic oscillators.**

Introduction

The general function of optical devices consists of the modification of the wavefront of light by altering its phase, amplitude, and polarization in a desired manner. The class of optical components with a varying phase retardation includes lenses, wave plates, spiral phase plates [1], axicons [2], and more generally spatial light modulators (SLMs), which are able to imitate many of these components by means of a dynamically tunable spatial response [3]. All of these conventional optical components rely on gradual evolution of phase, amplitude, and polarization as the wave propagates through an optically-thick medium. The introduction of abrupt phase changes into the optical path by using the resonant behavior of plasmonic nanostructures allows one to achieve control over the wavefront without relying on gradual phase accumulation [4]. This approach is now enabling the design of various new optical devices which are thin compared to the wavelength of light [5, 6, 7].

Our previous work on phase discontinuities involved spatially-inhomogeneous configurations of V-shaped optical antennas [4, 5, 7]. Here, we report that homogeneous arrays optical antennas supporting two independent and orthogonally-oriented current modes operate as highly birefringent meta-surfaces. We consider arrays of V-shaped antennas, creating a connection with our previous work, and Y-shaped antennas in which the anisotropy can be widely tailored or extinguished via interference between the scattered light from the two current modes. A simple, analytical two-oscillator model for two-dimensional (2D) optical antennas is developed which captures the physics of these antennas and provides an intuitive way to understand how engineering of the amplitude and phase of the scattered light provides control over the optical anisotropy of the meta-surface.

**Two-dimensional optical antennas as double oscillators**

The optical response of surface plasmons in confined structures such as linear optical antennas is well described in terms of a charge-oscillator model that includes the effects of forcing by the incident electric field, internal damping, and radiation reaction [8]. Of particular interest for this study are 2D optical antennas because they support orthogonal, independent current modes. By controlling their dimension and shape, their inherent optical anisotropy can be tailored over a broad range. In this section, we develop an extension of the charge-oscillator model for V-shaped optical antennas, which can serve as elements of birefringent meta-surfaces. Examples of other, previously investigated 2D plasmonic structures which support orthogonal modes include asymmetric cross antennas [9], L-shaped nanoparticles and antennas [10, 11], split rings [12], and rectangular patch antennas [13].

Treating the two plasmonic modes as independent harmonic oscillators, the combined system can be represented as a charged mass on two orthogonally-oriented springs (schematics in Fig. 1(a, b)). The two oscillators are oriented along $x$ and along $y$, respectively, with the incident light propagating along $z$ and its electric field oriented along an axis $w$, which lies in the $x$-$y$ plane at an angle $\theta$ from the y-axis (Fig. 1(c)). For a charged, driven oscillator oriented along the $x$-axis, the complex amplitude $x(\omega)$ of the displacement from the equilibrium position, assumed to be harmonically varying as $x(\omega)\, e^{i\omega t}$, can be written as [8]:

$$x(\omega) = \frac{\frac{q_x}{m_x}E_{0,x}}{(\omega_{0,x}^2 - \omega^2) + i\frac{\omega}{m_x}(\Gamma_{a,x} + \omega^2 \Gamma_{s,x})} = \tilde{x}(\omega)E_{0,x} \qquad (1)$$

where $q_x$ is the participating charge, $m_x$ is the mass, $\omega_{0,x}$ is the resonant frequency, and $\Gamma_{a,x}$ and $\Gamma_{s,x}$ are the damping coefficients representing absorption and scattering, respectively. The field emitted by the oscillator $E_{s,x}(\omega)$ can be written as

$$E_{s,x}(\omega) = -D_x(\mathbf{r})\sqrt{\Gamma_{s,x}}\,\omega^2 x(\omega). \qquad (2)$$

In Eqn. 2, $D_x(\mathbf{r})$ contains the angular and radial dependence of the emitted field. The exact form of $D_x(\mathbf{r})$ depends on the specific geometry of the oscillator and the surrounding environment, but in the limit that the oscillator element is small relative to the wavelength of light, $D_x(\mathbf{r})$ is simply the emission pattern of a radiating electric dipole [see Supplementary Information]. The expressions for the y-oscillator ($y(\omega)$, $\tilde{y}(\omega)$ and $E_{s,y}(\omega)$) are analogous to that of the x-oscillator.

In general, light is scattered by the two-oscillator element into some elliptical polarization. Deliberate engineering of this polarization state offers intriguing prospects, but remains outside of the scope of this work. Instead, we focus on light scattered into the polarization state along the v-axis in Fig. 1(c), which is the cross-polarized direction relative to the incident light. The polarization conversion efficiency is a direct measure of the degree of anisotropy of the 2D plasmonic antenna, and can be easily isolated by filtering out light with the incident polarization with a linear polarizer. This cross-polarized configuration is also critical for the design of planar optical components, as it extends the phase coverage of the scatterers to cover a full $2\pi$ range [4].

Given an incident field polarized along $\hat{w}$, (Fig. 1(c)), we calculate the component of the emitted field polarized along the v-direction $E_{s,v}(\omega)$. We can break up this polarization-conversion process into two steps: the in-coupling of incident light into the two oscillator modes, and the out-coupling of cross-polarized light. The in-coupling process involves the projection of the incident field along the two oscillator modes, which can be expressed as $E_{0,x} = E_0\hat{w}\cdot\hat{x} = E_0\sin(\theta)$ and $E_{0,y} = E_0\hat{w}\cdot\hat{y} = E_0\cos(\theta)$. For the out-coupling process, we project the field scattered by each oscillator onto the v-axis to arrive at

$$E_{s,x}\hat{v}\cdot\hat{x} = -D_x(\mathbf{r})\sqrt{\Gamma_{s,x}}\omega^2\tilde{x}(\omega)E_{0,x}\cos(\theta) \qquad (3)$$

$$E_{s,y}\hat{v}\cdot\hat{y} = D_y(\mathbf{r})\sqrt{\Gamma_{s,y}}\omega^2\tilde{y}(\omega)E_{0,y}\sin(\theta) \qquad (4)$$

After summing these projections, the total cross-polarized field emitted by the structure $E_{s,v}$ can be written as

$$E_{s,v}(\omega) = D(\mathbf{r})\frac{E_0}{2}\sin(2\theta)\omega^2\left[\sqrt{\Gamma_{s,x}}\tilde{x}(\omega)e^{i\pi} + \sqrt{\Gamma_{s,y}}\tilde{y}(\omega)\right] \qquad (5)$$

where we assumed that $D_x(\mathbf{r}) \approx D_y(\mathbf{r}) = D(\mathbf{r})$, which is true for light emitted approximately normal to the orientation of the two oscillators. Eqn. (5) provides a complete description of the generation of cross-polarized light by our two-oscillator system. The intensity $|E_{s,v}(\omega)|^2$ and phase $\phi(\omega)$ of the cross-polarized light ($E_{s,v}(\omega) = |E_{s,v}(\omega)|e^{i\phi(\omega)}$) are plotted in Fig. 1(d, e). The specific parameters $\Gamma_{a,i}, \Gamma_{s,i}, m_i, \omega_{0,i}$ ($i \in x, y$) for the two oscillators used in generating Fig. 1(d, e) correspond roughly to the modes of a typical isolated V-shaped antenna [see Supplementary Information].

The phase of the cross-polarized light generated by our two-oscillator element (black curve in Fig. 1(e)) is able to span twice the range of phase of either single oscillator (blue or pink), even though the two oscillators are uncoupled and operate independently. This phase extension, which can be seen as the $e^{i\pi}$ term in Eqn. (5), is due to the fact that the projections of the scattered fields from the spatially-overlapped x- and y-oriented oscillators onto the v-axis are opposite in phase (Fig. 1(c) and Eqns. (3) and (4)). This is shown graphically in Fig. 1(e) as a shift of the intrinsic phase response of the x-oscillator (blue curve) down by π. As a result, this two-oscillator system is able to provide a much larger phase coverage in cross-polarization while maintaining a significant scattering amplitude. The use of two spectrally-separate resonances as in Fig. 1 allows one to broaden the frequency range over which there is significant polarization conversion (Fig. 1(d)), creating broadband optical anisotropy.

Eqn. (5) encodes the $\theta$-dependence of the polarization conversion properties with the $\sin(2\theta)$ term. No cross-polarized light is generated for $\theta = 0°$ or $90°$ when the incident field is aligned along one of the two orthogonally-oriented oscillators, and maximum polarization conversion is obtained for $\theta = 45°$. Due to the $\sin(2\theta)$ dependence, a rotation of the structure by 90° maintains the amplitude of cross-polarized scattering while adding an extra phase of π to the scattered light. This feature of double-oscillators allows for the same element to scatter with phase $\varphi$ and $\varphi + \pi$, simply by applying a 90° rotation $(\theta' = \theta + 90°$ in Eqn. (5)), and was used in [4] to generate 8 distinct phase elements from four V-shaped antennas.

The intensity and phase response of cross-polarized scattered light in Fig. 1(d, e) is given as a function of wavelength (or, equivalently, frequency) for a fixed set of oscillator parameters. However, a careful analysis of Eqn. (1) leads to an alternative approach; since Eqn. (1) depends on $(\omega_0^2 - \omega^2)$, one can plot the amplitude and phase response of an oscillator as a function of $\omega_0$ for a fixed frequency $\omega$. This method of analyzing the behavior of oscillators is required for the design of phase elements for single-frequency optical components [4, 5, 14]. These two approaches are complementary; if wide tunability of the phase response for a fixed oscillator (or set of oscillators) is achieved as a function of frequency, then it can likewise be achieved for a fixed operating frequency by exploring the parameter space of the oscillator.

**Broadband spectral and polarization-conversion properties of V-antenna arrays**

To experimentally characterize the birefringent properties of meta-surfaces based on V-shaped antennas, we performed mid-IR spectral measurements of arrays fabricated by electron-beam lithography (see SEM in Fig. 2) using a Fourier transform infrared (FTIR) spectrometer. In Fig. 2(a, b) we plot the measured (1 - $T$) spectra, where $T$ is the transmission through arrays of V-shaped gold antennas ($L \approx 650$ nm, $w \approx 130$ nm, Si substrate) for opening angles $\Delta$ from 45° to 180° at normal incidence. The quantity (1 - $T$) corresponds to the sum of the scattering and absorption spectra, and is sometimes referred to as extinction. The incident polarization was fixed along the x- and y- directions in (a) and (b), respectively.

The (1 - $T$) spectra peak around the resonance frequencies of the antenna modes, and the expected small differences between the spectral locations of the scattering and absorption peaks [8] are obscured by the inhomogeneities of the arrays. The first order approximation of the locations of the two peaks yields $\lambda_{0,x} \simeq 2Ln_{eff} = 3.4 \mu m$ and $\lambda_{0,y} \simeq 4Ln_{eff} = 6.8 \mu m$, taking $n_{eff}$ as 2.6 [4]. The measurements yield slightly different values ($\lambda_{0,x} \simeq 3.7 \mu m$ and $\lambda_{0,y} \simeq 6 \mu m$ for $\Delta = 90°$, e.g.) due to the non-zero thickness and width of the antennas, coupling between neighboring elements, and the presence of native oxide on the silicon surface (~2 nm as measured by ellipsometry), which is also responsible for the sharp feature at ~8 - 8.5 μm [15, 16] [See supplementary information]. The spectral position of the resonances shifts with varying Δ due to near-field interactions between the two arms, which are strongest for small Δ. In Fig. 2(c), we plotted the power of the generated beam in cross-polarization, normalized to the incident light, when the polarization is along $\theta$ = 45°, where $\theta$ is defined in Fig. 1(c). As expected, the polarization conversion efficiency peaks in the 3 μm – 8 μm range, in the vicinity of the antenna resonances. In Fig. 2(d), we plotted the line scans of Figs. 2(a - c) for Δ = 90°, indicated by the white dashed lines. The small, irregular oscillations between ~4.5 μm and ~7.5 μm correspond to atmospheric absorption from ambient water vapor. The corresponding finite-difference time-domain (FDTD) simulations are shown in Fig. 2(e). Note that our choice of Δ = 90° is purely arbitrary and one can generate similar plots for any opening angle. Additionally note that in the experiment a finite, 280 μm thick double-side polished silicon substrate is used; however, it is computationally intensive to include this large, finite slab in the simulations due to the high resolution required to model the nanoscale antennas. The polarization-conversion result shown as the black curve in Fig. 2(e) is a result of such a complete simulation and matches very well the experimental measurements, but all of the other simulations presented in this article are performed using an infinite silicon substrate. [see Supplementary Information for more details, as well as for simulations corresponding to Fig. 2(a-c)].

In Fig. 2(f), we plot the measured polarization-conversion efficiency at λ = 4μm (wavelength chosen to be away from the atmospheric absorption resonances) as a function of $\theta$. The error bars account for polarizer misalignment Δ$\theta$ and spectrometer noise. The data was fit to $A\sin^2(2\theta)+C$ to account for the $\theta$-dependence of Eqn. (5) and some offset due to imperfect cross-polarization extinction of our polarizers (black curve). Since within the 0 – 90° range, the $\theta$-dependence only affects the amplitude of $E_{s,v}$, $\theta$ can be used as a degree of freedom to control the cross-polarized scattering amplitude without altering its phase response (for example, simply rotating the individual elements of a planar optical component such as the vortex plates in ref. [5] allows for independent amplitude control of each antenna, enabling the creation of a simultaneous amplitude and phase plate). The phase of the cross-polarized light (which is independent of $\theta$) is calculated via FDTD simulations, and plotted as the black curve in Fig. 2(g). As in Fig. 1(e), the brightness of the curve encodes the intensity of the scattered light.

While the intensity and phase response in Fig. 2(d, e, g) is given as a function of wavelength, a complementary plot can be made keeping the wavelength constant and sweeping across the arm length $L$ because the resonant wavelengths of both oscillator modes $\lambda_{0,x}$ and $\lambda_{0,y}$ depend linearly on $L$, though an extra scaling factor has to be introduced in the near-IR and visible regimes [17]. It is much more difficult to generate such plots because a new sample has to be fabricated and measured (or simulated) for every possible value of $L$; however, the Supplementary Information includes approximate intensity and phase response maps for V-antennas as a function of geometrical parameters $L$ and $\Delta$, keeping the frequency constant.

**Tailorable optical anisotropy via interference in Y-shaped antennas**

The spectral position of V-antenna resonances can be tuned by varying the arm length L and, to a smaller extent, by adjusting the opening angle $\Delta$ (Fig. 1(a, c)). However, both of these simultaneously shift the resonance frequencies of the x-oriented (symmetric) and y-oriented (antisymmetric) modes of the antenna. By appending a "tail" of length $L_T$ to the V-antenna as shown in Fig. 3(a), an additional degree of freedom is attained that allows for independent tuning of the spectral position of the x-oriented mode. By increasing $L_T$, the x-oriented mode is red-shifted without affecting the y-oriented mode. We fabricated these Y-shaped antennas on a silicon substrate, with an SEM image of the structures shown in Fig. 3(a). The x- and y-oriented modes are identified in Fig. 3(b, c), respectively, for 4 different values of $L_T$, by measuring the reflectivity spectra from arrays of these antennas. The x-oriented mode increases in resonant wavelength and amplitude as $L_T$ increases, because the effective antenna length is increasing, increasing $q_x$, $m_x$, and $\Gamma_{s,x}$ and decreasing $\omega_{0,x}$ (Fig. 3(b)). The y-oriented mode is not perturbed by this tail section, so all of the reflectivity curves in Fig. 3(c) overlap.

The polarization conversion efficiency due to the Y-antennas is plotted in Fig. 3(d) as a function of wavelength, given incident polarization along $E_0 \hat{w}$ for $\theta = 45°$ as in Figs. 1 and 2, such that the projections of the incident field along the two antenna modes are equal, which maximizes the conversion. There is a substantial amount of polarization conversion for $L_T \simeq 100nm$, $300nm$, and $700nm$ (red, black, and blue curves, respectively). However, for $L_T \simeq 500nm$ (green curve), the polarization conversion is almost completely extinguished. FDTD simulations corresponding to Fig. 3(b-d) are shown in Fig. 3(f-h), and demonstrate the same behavior as in the measurements (unlike the simulations in Fig. 2(b) and (d), we did not include the native oxide layer in the simulations as it detracts from visual clarity, so the feature at λ ~ 8 μm is not reproduced in these simulation results). The origin of this effect can be interpreted as destructive interference between the contributions to the cross-polarization generation from the two oscillator modes. As illustrated in Fig. 3(e), the incident field excites both the x-oriented and y-oriented modes of the Y-antenna, each of which contribute to the cross-polarized field. However the projections of the scattering of the x- and y-oriented oscillators onto the v-axis are opposite in phase (Fig. 1(c)), which results in an additional dephasing of π between the two contributions ($e^{i\pi}$ term in Eqn. (5)).

When the two oscillators are nearly identical in their individual amplitude and phase response (as is the case for $L_T \simeq 500 nm$), their contributions to the polarization conversion efficiency are π out of phase, resulting in destructive interference [see Supplementary Information for a detailed discussion]. The observed imperfect extinction and line shape asymmetry are a result of the two eigenmodes in our experiment being not completely identical in linewidth, amplitude, and resonance frequency. In this way, Eqn. (5) explains that any structure with, for example, three-fold ($C_3$) or four-fold ($C_4$) rotational symmetry cannot be used for polarization conversion, and is therefore isotropic. Conversely, arrays of resonant structures which support two unequal eigenmodes can be viewed as meta-surfaces with giant birefringence since they can rotate the polarization of light over a thickness of just ~50 nm at mid-IR wavelengths. Such birefringence arising due to structural anisotropy instead of intrinsic crystal properties of a material is sometimes referred to as "form birefringence" in literature (see, e.g., [9, 11, 18-22]). Conversely, arrays of $C_3$-symmetric structures including Y-shaped antennas with equal arms (and Δ = 120°) and nanoparticle trimers [23] have an isotropic in-plane response [24], and thus have no polarization-converting properties. We note that resonant metallic structures which exhibit large optical activity have been explored in the literature (e.g. [25-26]), and serve as the circular-polarization equivalents to the birefringent metallic structures demonstrated in this work.

**Conclusion**

We showed that an optical element consisting of two orthogonally-oriented uncoupled oscillators possesses widely-tailorable optical anisotropy, and is able to independently control the phase and amplitude of light scattered into the cross-polarization. To demonstrate this concept, we utilized the double resonances of V- and Y-shaped plasmonic antennas. The properties of these resonances are controlled by changing the opening angle and length of the arms of both the V- and Y-shaped antennas. While the Y-shaped antennas have a larger geometrical footprint than their V counterparts, they allow for more direct tailoring of the two eigenmodes and thus are more flexible components of anisotropic interfaces. The phase response of the individual modes in Y-shaped antennas was experimentally investigated via an interference experiment, in which the polarization conversion was extinguished by destructive interference between the contributions from two nearly-identical oscillator modes. Arrays of these antennas form meta-surfaces with widely tailorable birefringence.

The approach of molding waves with both uniform and spatially-varying arrays of antennas is applicable to a large part of the electromagnetic spectrum from radio frequencies to the ultraviolet. The possibility of using other optical oscillators, such as quantum dots, nanocrystals, resonant molecules or metamolecules is promising for creating isotropic or birefringent optical phase and amplitude elements with deeply-subwavelength dimensions, reduced losses and dynamic tunability.

**Methods**

Fabrication

The antenna arrays were fabricated on high resistivity (> 10,000 Ω-cm) double side polished silicon using a conventional electron-beam (e-beam) lithography process with lift-off. A double layer of poly(methyl methacrylate) (PMMA) resist (495A4, then 950A2, MicroChem) was spun at 4000 RPM onto the silicon wafer, baked at 120°C, and then exposed using a 100 kV e-beam system (Elionix ELS-7000). After development with 3:1 isopropanol (IPA) : methyl isobutyl ketone (MIBK), 10 nm of titanium and 40 nm of gold was deposited using e-beam evaporation, and the lift-off process was completed in acetone with ultrasonic agitation. For SEM images of the resulting structures, see Fig. 2(a), in Fig. 3(a) as well as the Supplementary Information.

Measurements

The transmission and polarization conversion measurements were performed using a Bruker Vertex 70 FTIR spectrometer connected to a Hyperion 2000 mid-IR microscope. A linearly polarized Globar source is used, and the beam is focused onto the sample and then collected in transmission mode using two 15X objectives (NA = 0.4). An additional polarizer was placed between the sample and the outcoupling objective for the polarization conversion measurements and is used to isolate the scattered light into the cross-polarization. For transmission measurements, the measured signal through V- and Y-antennas on a Si substrate was normalized to the transmission through the Si substrate alone. For polarization conversion measurements, the normalization reference was taken by doing a transmission measurement through the bare Si substrate with both polarizers aligned. The uncertainty in the measurement of the spectrum away from atmospheric noise was ~3%, and the uncertainty in θ (polarizer alignment) was ~2°. These uncertainties were combined using standard error propagation to obtain the error bars in Fig. 2(c).

Simulations

The FDTD simulations were performed using Lumerical Solutions FDTD running on a workstation. For Fig. 2(b, d), the transmission and cross-polarization properties of V-antenna arrays were simulated by using periodic boundary conditions to define a single repeating unit cell of (1 μm x 1 μm), each containing a single antenna. A broadband plane wave was launched from the silicon side to illuminate the antenna array. The phase response in Fig. 2(d) was obtained by performing a near- to far-field transform and looking at a point 1 meter above the antenna array in a direction normal to the surface. The phase response was calculated as the phase of the field at that point less the phase accumulated by the plane wave via propagation through the silicon slab and the air above the array. For Fig. 3(f-h), the simulations of the Y-antennas were done by defining a 3 x 3 array of antennas at a spacing of 1.7 μm x 1.7 μm, and using a total-field scattered-field (TFSF) plane wave source with perfectly matched layer (PML) boundary conditions [See Supplementary Information for more details].

**Acknowledgements**

The authors acknowledge helpful discussions with Jonathan Fan, and support from the National Science Foundation, Harvard Nanoscale Science and Engineering Center (NSEC) under contract NSF/PHY 06-46094, and the Center for Nanoscale Systems (CNS) at Harvard University. This work was supported in part by the Defense Advanced Research Projects Agency (DARPA) N/MEMS S&T Fundamentals program under grant N66001-10-1-4008 issued by the Space and Naval Warfare Systems Center Pacific (SPAWAR). M. Kats is supported by the National Science Foundation through a Graduate Research Fellowship. P. Genevet acknowledges support from the Robert A. Welch Foundation (A-1261). Z. Gaburro acknowledges funding from the European Communities Seventh Framework Programme (FP7/2007-2013) under grant agreement PIOF-GA-2009-235860. Harvard CNS is a member of the National Nanotechnology Infrastructure Network (NNIN).

**Figures**

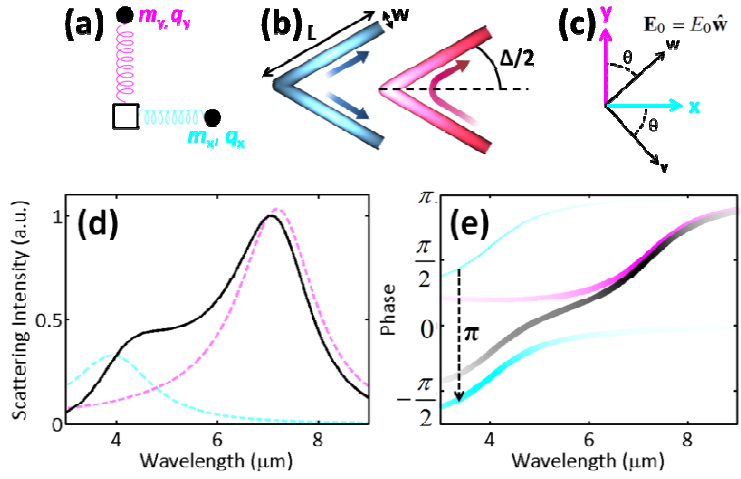

Fig. 1 (a) Charge-oscillator model of a two-mode plasmonic element, where $q$ is the charge and $m$ is the inertial mass. (b) A metallic V-shaped antenna which supports two orthogonal charge-oscillation modes (blue and pink). The arrows show the direction of current flow, and the colors represent the current density (lighter color indicates higher current). The axis of symmetry is marked with a dashed line. (c) Two coordinate systems related by a rotation by angle $\theta$. The $x$-$y$ axes are along the two fundamental oscillator modes, the $w$-axis is along the polarization of the incident field $E_0$, and the v-axis is along the cross-polarized component of the emitted field. (d) Calculated intensity $|E|^2$ of the field scattered into the cross-polarization due to two individual oscillators representing, in this case, two modes of a V-shaped antenna (blue and pink) with $\Delta = 90°$ and $L = 650$ nm, and due to the combined system (black) for $\theta = 45°$. Note that the black curve is not simply the sum of the blue and pink curves because of the coherent addition of fields. (e) Phase of the field scattered into the cross-polarization by the individual oscillators (blue and pink), and by the combined system (black). The brightness of the color encodes the scattering intensity, with darker colors signifying more intense scattering. The blue curve (due to the $x$-oscillator) is shifted down by $\pi$ from its intrinsic oscillator phase due to the geometry (see last term in Eqn. (5)).

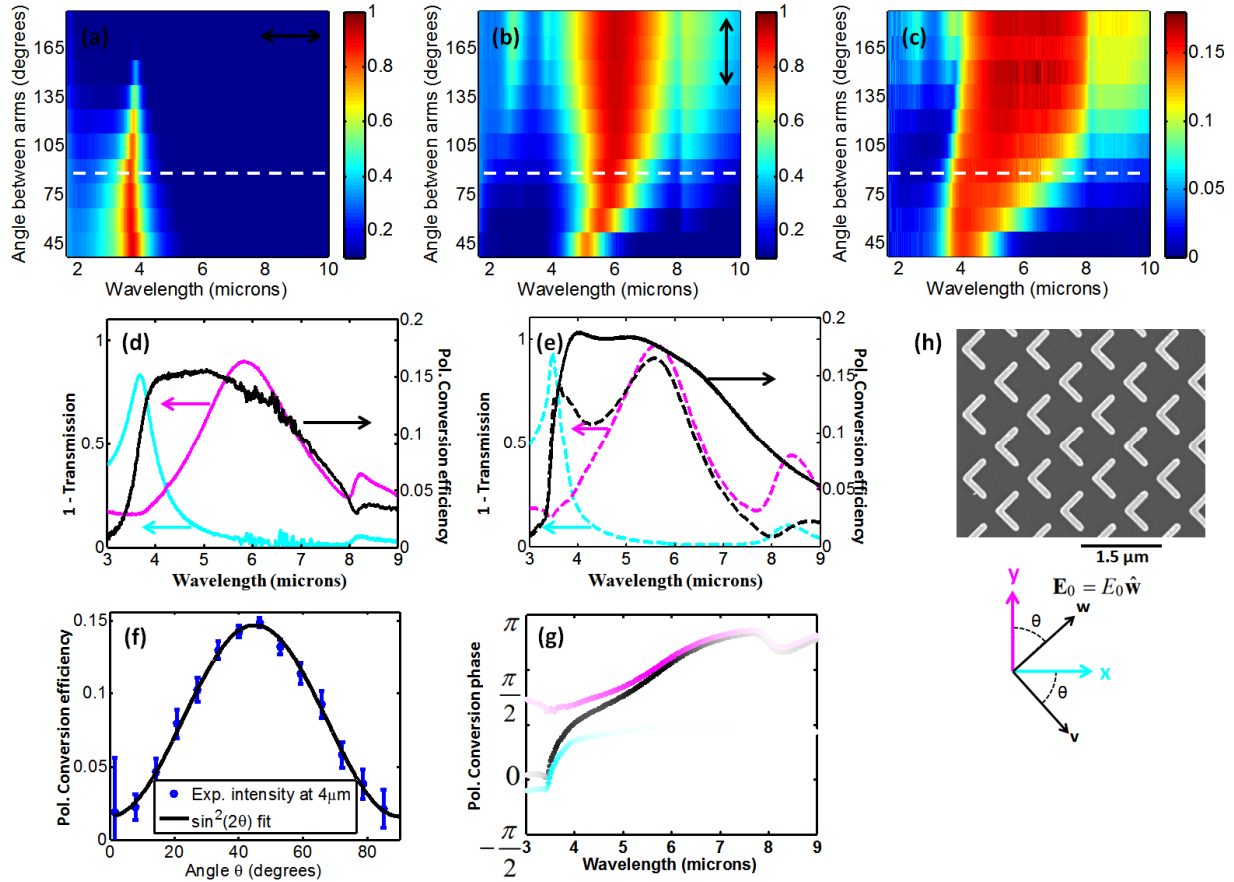

Fig. 2 (a, b). Color maps showing the measured extinction (defined as 1 - Transmission) spectra through arrays of lithographically-defined V-antennas ($L \sim 650$ nm, $\Delta$ from 45° to 180°) for $x$- and $y$-oriented incident polarizations, respectively. (c). Color map showing the polarization conversion efficiency spectra corresponding to (a, b) when the incident field is polarized at a $\theta = 45°$ angle between both principle axes. (d). Blue and pink curves are the measured extinction through the V-antenna array with $\Delta = 90°$ for the x-oriented (blue) and y-oriented (pink) incident polarization for every antenna. The black curve is the polarization conversion efficiency from the array for $\theta = 45°$. All three curves correspond to line scans of (a-c), shown by the white dashed line. (e). FDTD simulations corresponding to the curves in (d), with the dashed curves representing simulations with an infinitely thick substrate and the solid black curve representing the calculated polarization-conversion efficiency when the finite (280 µm) thickness of the substrate is accounted for. (f). Measured polarization conversion efficiency at $\lambda = 4$ µm plotted vs. the incident polarization angle $\theta$ (blue symbols) and a fit to a $\sin^2(2\theta)$ dependence as predicted by Eqn. (5). The calculated correlation $R$ is 0.997. (g). Phase response of the cross-polarized light generated by the antennas as calculated by FDTD (black). The blue and pink curves represent the phases of the contributions from the symmetric and antisymmetric modes, respectively. Brightness of the phase curves indicates the intensity of the scattered light. (h). SEM image of the $\Delta = 90°$ V-antenna array and the coordinate system.

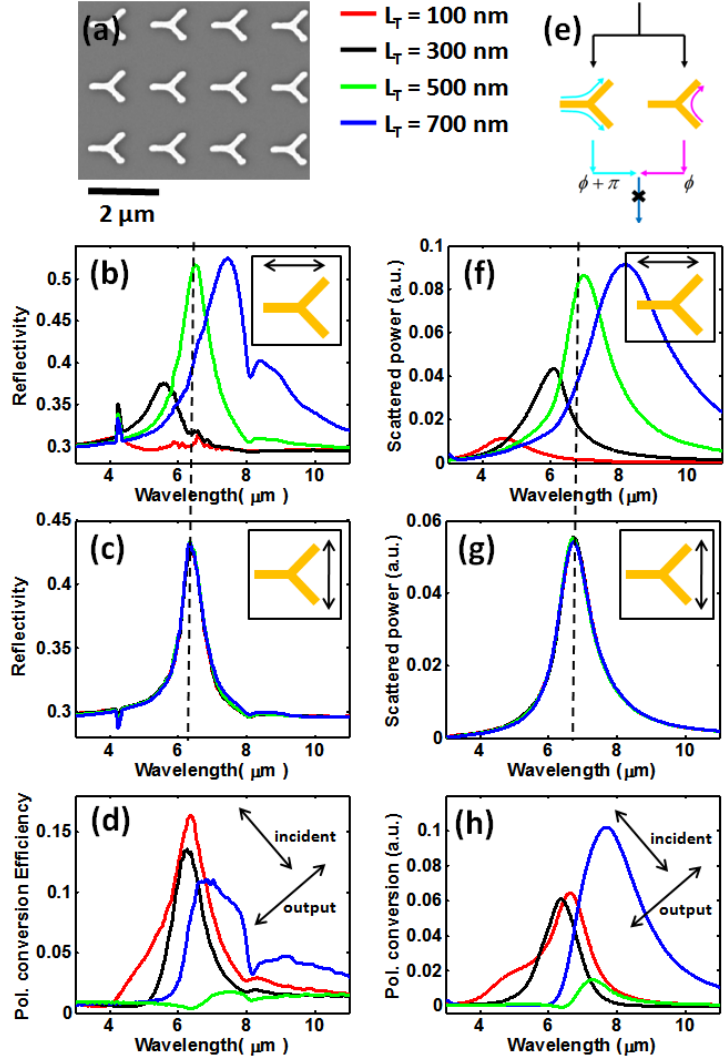

**Fig. 3** Y-shaped plasmonic antennas. (a) SEM image of the antenna array. (b, c) Measured normal-incidence reflectivity spectra of the x- and y-oriented antenna modes, respectively, as a function of tail length $L_T$, which varies from 100 nm to 700 nm by increments of 200 nm (see legend in upper panel). The reflectivity of the bare silicon substrate is ~0.3. The vertical dashed line shows that for $L_T \simeq 500 nm$ (green curves), the x- and y- oriented resonances are overlapping. The arrows indicate the polarization of the incident field. (d) Polarization conversion spectrum with θ = 45°, with the incident and measured polarizations indicated with arrows. The polarization conversion is nearly extinguished for one intermediate value of $L_T$ (green). (e) Diagram explaining the extinguishing of polarization conversion due of destructive interference between contributions of the two antenna modes when $L_T$ is adjusted such that the two oscillator modes have the same resonant response. (f-h) FDTD simulations corresponding to the measurements in (b-d). The native oxide layer was not included in the simulations for visual clarity.

**Supplementary information**

**Giant birefringence in optical antenna arrays with widely tailorable optical anisotropy**

**Mikhail A. Kats[1], Patrice Genevet[1,2], Guillaume Aoust[1,3], Nanfang Yu[1], Romain Blanchard[1], Francesco Aieta[1,4], Zeno Gaburro[1,5], and Federico Capasso[1]**

[1]School of Engineering and Applied Sciences, Harvard University, Cambridge, Massachusetts 02138, USA

[2]Institute for Quantum Studies and Department of Physics, Texas A&M University, College Station, Texas 77843, USA

[3]Department of Physics, Ecole Polytechnique, Paris, France

[4]Dipartimento di Fisica e Ingegneria dei Materiali e del Territorio, Università Politecnica delle Marche, via Brecce Bianche, 60131 Ancona, Italy

[5]Dipartimento di Fisica, Università degli Studi di Trento, via Sommarive 14, 38100 Trento, Italy

**SEM images of fabricated structures**

Scanning electron microscope (SEM) images of the V- and Y-shaped antennas we fabricated are shown in the inset of Fig. 2(a), in Fig. 3(a), and in Fig. S1.

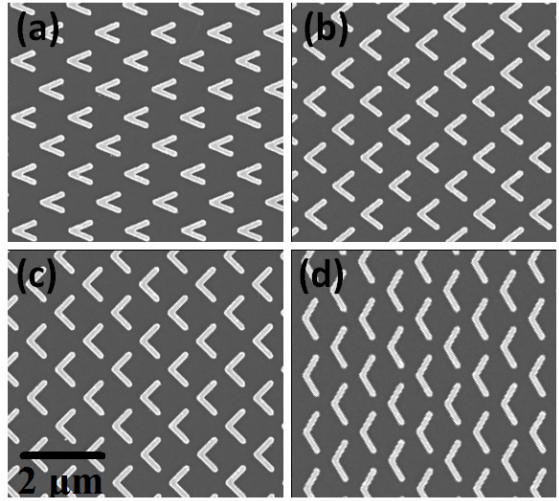

**Fig. S1.** SEM images of gold V-shaped antennas fabricated on a silicon substrate with opening angle (a) $\Delta = 45°$, (b) $\Delta = 75°$, (c) $\Delta = 90°$, and (d) $\Delta = 120°$

**Simulations of V-antenna arrays**

In Fig. S2, we map the two modes of the V-antennas in the antenna arrays as a function of both wavelength and changing angle $\Delta$ from 45° to 180° by showing the measured (a-c) and calculated (d-f) transmission spectra. The (a) and (b) panels correspond go Fig. 2(a, b) in the text.

The orientation of the incident polarization is shown in the upper right corner. Fig. S2 (a) and (d) correspond to excitation of only the x-oriented symmetric antenna mode, whereas (b) and (e) correspond to the y-oriented antisymmetric mode, and (c) and (f) shows both excited modes. All of the experimental spectra are reproduced very well in simulations, including the feature at 8-9μm due to a phonon resonance in the native silicon oxide layer which is enhanced by the strong local fields formed around the metallic antennas. In Fig. S2 (b) and (e), a higher order antenna mode is clearly visible at $\lambda_0 \sim 3$ μm for large $\Delta$.

The resonant properties of the two modes of V-shaped plasmonic antennas (Fig. 1(b)) are associated with the length L of one of the antenna arms, and the combined length 2L, respectively. Changing the angle $\Delta$ at which the arms are oriented changes the radiative losses $\Gamma_{s,i}$ of the oscillators $(i \in \{x, y\})$, with $\Gamma_{s,x}$ increasing as $\Delta$ decreases due to the greater overlap of the incident field with the x-oscillator and $\Gamma_{s,y}$ decreasing due to the reduced overlap with the y-oscillator.

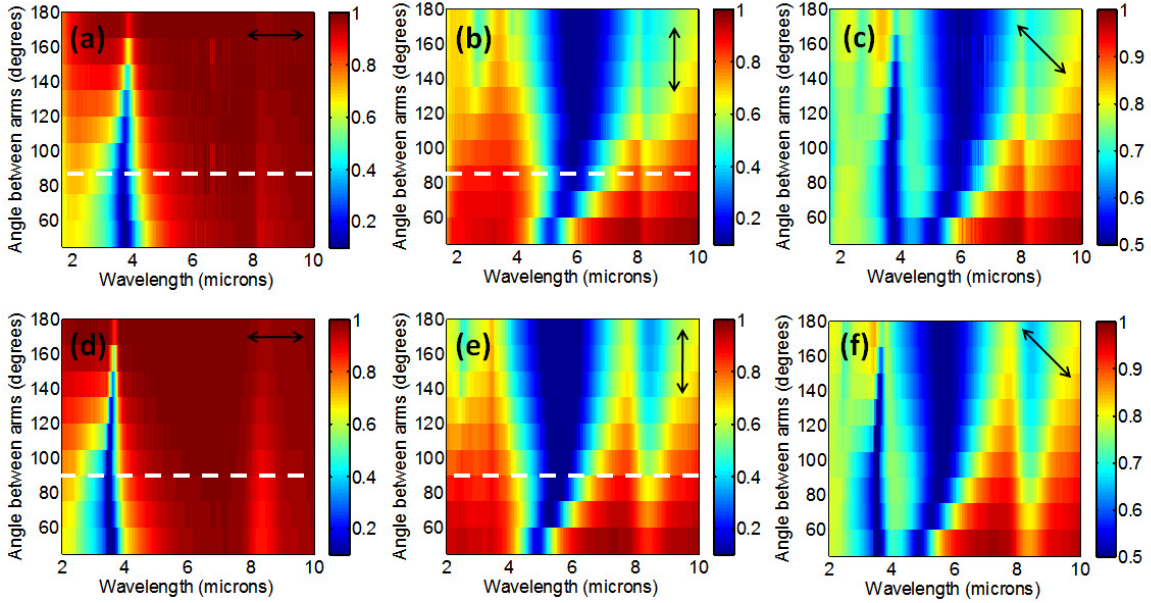

**Fig. S2** Mapping the two V-antenna eigenmodes corresponding to the two-oscillators. (a-c) Measured transmission spectrum through the V-antenna arrays at normal incidence as a function of wavelength (horizontal axis) and angle $\Delta$ (vertical axis) for fixed arm length L = 650 nm. The incident light is polarized (a) along the symmetry axis of the antennas, (b) orthogonal to the symmetry axis, and (c) at a 45° angle. (d-f) FDTD simulations corresponding to the experimental spectra in (a-c), respectively. The feature at $\lambda$ = 8 - 9 µm is due to the phonon resonance in the ~2 nm native oxide. The horizontal lines represent the line scans taken to obtain Fig. 2 in the main text.

We re-plotted the polarization conversion spectral map from Fig. 2(c) in Fig. S3(a). The corresponding FDTD simulation is shown in Fig. S3(b), and retains the same features as the experiment, though the simulated polarization conversion spectrum is more clearly broken up into two resonances. The experimental data shows less of this separation due to inhomogeneous broadening in the experiment due to fabrication imperfections. The simulated phase of the scattered light is plotted in Fig. S3(c).

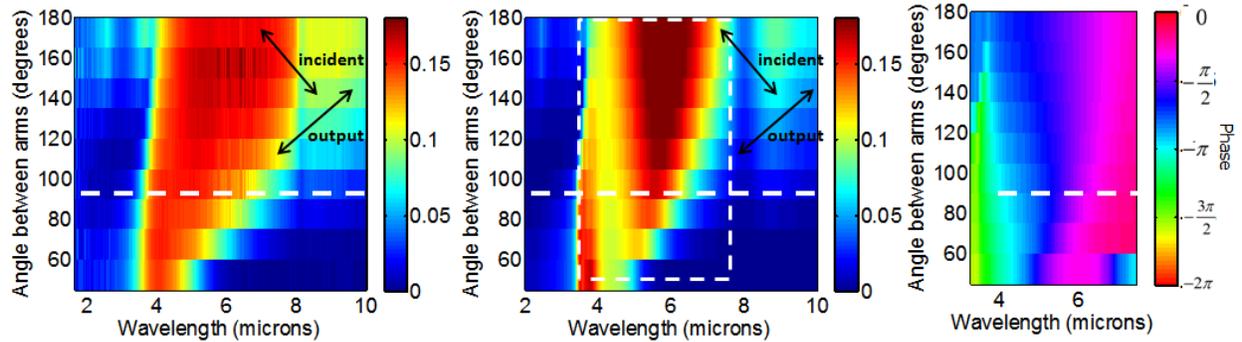

**Fig. S3** (a,b) Experimental measurements (a) and FDTD calculations (b) of the cross-polarized scattering for the V-antenna arrays in Fig. 2. The arrows indicated the polarization of the incident and output (collected) light. (c) Calculated phase of the cross-polarized light where in the spectral region where scattering efficiency is high (box enclosed by dashed lines). The horizontal lines represent the line scans taken to obtain Fig. 2 in the main text.

**Comment on the simulations of Y-antennas**

The inter-antenna spacing was increased due to the increased geometric footprint of the Y-antennas relative to the V-antennas, and the 3 x 3 array was used instead of the infinite array to avoid observing a Rayleigh-Wood anomaly feature [S1, S2], which is not present in our experiment due to the finite size of our arrays, inhomogeneous broadening effects, and the finite size of our beam. We previously performed such small-array simulations of plasmonic arrays in [S3].

**The oscillator models of plasmonic antennas**

As shown in [7], an optical antenna (or any localized plasmonic resonance) can be treated as a driven harmonic oscillator with two damping terms -- the first due to internal absorption via free carrier absorption in the metal (as well as any other optical losses due to the surrounding dielectric), and the second due to emission of light into free space (scattering). In the case of this paper, our elements contain two independent, orthogonal plasmonic modes, each of which we treat as a damped, driven harmonic oscillator. In this section, we re-state and elaborate on some of the results from [7] that are relevant to this work, leaving off the x- and y- subscripts because the x- and y- oscillators can be treated independently, and using x as the position variable.

A single plasmonic mode can be modeled as a charge q located at x(t) with mass m on a spring with spring constant $\kappa$. The charge q and mass m are, roughly speaking, the charge and the mass of the conduction electrons which interact with external driving and scattered fields. The oscillator equation can be written as

$$m\frac{d^2x}{dt^2} + \Gamma_a \frac{dx}{dt} + \kappa x = qE_0 e^{i\omega t} + \Gamma_s \frac{d^3x}{dt^3}$$

with internal damping coefficient $\Gamma_a$ and radiation damping coefficient $\Gamma_s = q^2/6\pi\epsilon_0 c^3$. The time-harmonic solution of this equation (with the $e^{i\omega t}$ term omitted) is

$$x(\omega) = \frac{(q/m)E_0}{(\omega_0^2 - \omega^2) + i\frac{\omega}{m}(\Gamma_a + \omega^2 \Gamma_s)}$$

where $\omega_0 = \sqrt{k/m}$. The time-average scattered power $P_s(\omega)$ can be written as

$$P_s(\omega) = \omega^4 \Gamma_s |x(\omega)|^2.$$

The scattered electric field from the oscillator is polarized along the x-direction and can be written as

$$E_s(\omega) = -D(\mathbf{r})\sqrt{\Gamma_s}\,\omega^2 x(\omega).$$

$D(\mathbf{r})$ contains the angular and radial dependence of the emitted field, which changes significantly with the geometry of the oscillator. If the oscillator is very small relative to the wavelength of light, $D(\mathbf{r})$ has the angular dependence of a radiating electric dipole. When located in vacuum, $D(\mathbf{r})$ is proportional to $\sin(\phi)/r$, where $\phi$ is the polar angle from the dipole axis and r is the distance from the dipole [S4]. However, as the oscillator size approaches the scale of the wavelength, $D(\mathbf{r})$ becomes significantly more complex. In a forthcoming paper [S5] we show that for V-shaped antennas such as the ones studied in Figs. 2 and 3 both the amplitude and phase response vary as a function of far-field angle, making $D(\mathbf{r})$ in general complex-valued. $D(\mathbf{r})$ follows the following:

$$P_s(\omega) = \Gamma_s \omega^4 |x(\omega)|^2 = \oiint_S |E_s(\omega)|^2 d\Omega = \oiint_S |D(\mathbf{r})\sqrt{\Gamma_s}\,\omega^2 x(\omega)|^2 d\Omega$$

where $d\Omega$ is the solid angle, and S is any surface which encompasses the scattering element.

Furthermore, in practical realizations of meta-surfaces based on plasmonic oscillators, the elements are typically fabricated on a substrate, such that they are surrounded by two half-spaces with differing dielectric properties. This further alters the emission pattern [S5, S6].

**Note regarding the charges $q_x$ and $q_y$**

In Eqn. (1), it is stated that $q_x$ is the total charge participating in the x-oriented oscillation behavior. Correspondingly, $q_y$ is charge involved in the y-oriented oscillation. Note that for any asymmetric plasmonic particles, $q_x$ will not in general be equal to $q_y$. This is best illustrated in the Y-antennas in Fig. 3(e). As the tail length $L_T$ is altered, more charges participate in the x-oriented oscillation, so $q_x$ changes, while $q_y$ remains unchanged.

**Obtaining parameters for Eqn. 1 by fitting to simulations**

In order to obtain the parameters used to generate Fig. 1 in the main text, we fit the scattering spectra of the two orthogonal antenna modes as calculated by FDTD simulations to Eqns. (1) and (2) in the main text. An isolated gold antenna with $\Delta = 90°$ and L = 650 nm and sitting on a silicon substrate is used in the scattering simulations. The resulting fits are shown in Fig. S4 for the symmetric mode (a) and the antisymmetric mode (b).

Since the internal losses of antennas at these wavelengths are approximately an order of magnitude smaller than the scattering losses [4], we assumed here that they were negligible, reducing the number of fitting parameters. The resulting values are $\omega_{0,x} = 5.18 \times 10^{14} s^{-1}$, $\omega_{0,y} = 2.67 \times 10^{14} s^{-1}$, $\Gamma_{a,x}/m_x = 0$, $\Gamma_{a,y}/m_y = 0$, $\Gamma_{s,x}/m_x = 13 \times 10^{-16} s$, $\Gamma_{s,y}/m_y = 10.4 \times 10^{-16} s$, with $(q_x/m_x)/(q_y/m_y) = 1.18$. Note that only relative values are given for the charge and mass parameters as the data in Figs. 1(d) and S4 is in arbitrary units.

The simulations (blue dots) and the resulting fit (black line) match up very well.

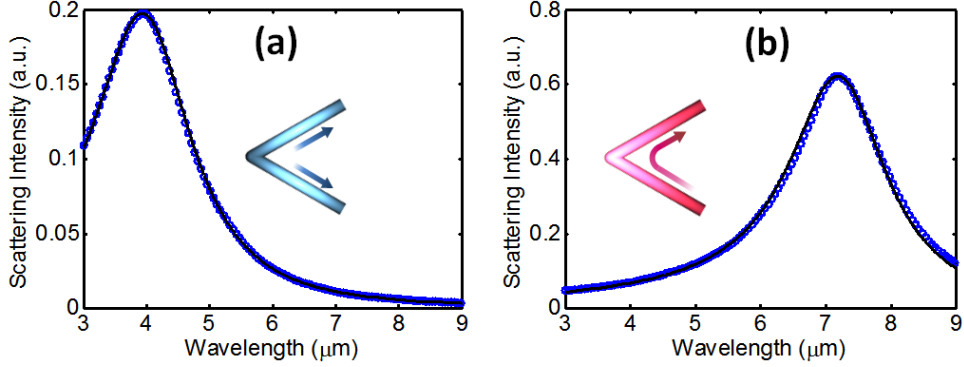

**Fig. S4** Scattering spectra from the symmetric (a) and antisymmetric (b) mode of an isolated gold V-shaped antenna (Δ = 90° and L = 650 nm) on a silicon substrate as calculated by FDTD simulations (blue dots). The black curve shows the fit of Eqns. (1) and (2) to the simulation results.

**Discussion on dispersion and Q-factors**

Fundamentally, the phase and amplitude control over the cross-polarized scattered light (and thus the optical anisotropy) offered by our two-oscillator system is due to the dispersive nature of the oscillators, corresponding to a significantly reduced speed of light. The geometry and type of oscillator can be used to obtain significant control over the dispersion and speed of light at the oscillators (e. g. [S7, S8]). The oscillators presented here have relatively low quality factors, corresponding to broad linewidths and relatively low dispersion. For the general design of phase and amplitude components with oscillator elements, the resonances must be sufficiently narrow to achieve a significant phase coverage yet sufficiently broad to enable broadband operation. Furthermore, more broad resonances allow for greater tolerances for practical device implementation. The relatively low Q-factors of optical antennas offer a good compromise.

**Additional Y-antenna measurements**

The measurements shown in Fig. 3(b-d) were also performed for Y-antennas with multiple angles Δ, although data for only one angle was presented in Fig. 3. In Fig. S5, we summarize the experimentally-measured resonant wavelengths of the x-oriented (circles, dashed line) and y-oriented (squares, solid line) oscillator modes for values of Δ from 60° to 120°. The extinction of polarization conversion due to destructive interference occurs whenever the two modes overlap in spectrum (i. e. when the solid and dashed lines cross).

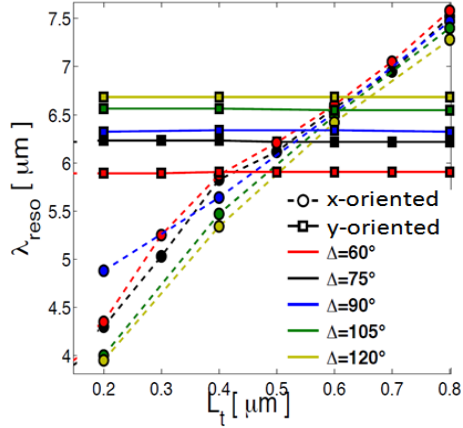

**Fig. S5** Experimental mapping of the resonance wavelength of the x-oriented and y-oriented oscillator modes of Y antennas for various opening angles Δ as a function of the tail length $L_T$. The resonance wavelength is calculated by taking the peak of the reflectance for each antenna geometry as a function of wavelength.

**Interference in polarization conversion**

In Section 3 of the main text, we made the claim that the extinguishing of polarization conversion in certain Y-shaped antennas is due to destructive interference between the two oscillator modes. In this supplementary section, we use the model developed in Section 1 to demonstrate how the polarization conversion efficiency evolves with various parameters for the two oscillator modes.

In Figs. S6 and S7 we plot the intensity and phases of light scattered into the cross-polarization by a two oscillator element following the same convention and coordinate systems as Figs. 1 and 2 in the main text. The blue and pink curves represent the polarization conversion intensity due to the x- and y-oriented modes, respectively, while the black curve represents the total polarization conversion intensity from the two-oscillator element, which includes the interference between the two independent contributions.

In Fig. S6(a), we plot the polarization conversion contributions from two identical orthogonal oscillators, with resonant wavelengths around 5 μm. Note that the blue and pink curves are overlapping exactly. The phase response of these two contributions to the polarization conversion is show in Fig. S6(d). There is a π phase difference between the two curves across all wavelengths; this is a graphical representation of the $e^{i\pi}$ term in Eqn. 5 and is due to the fact that the projections of the x- and y- oscillators onto the v-axis in Fig. 1 are exactly out of phase. As a result, the two contributions completely destructively interfere, and therefore the total polarization conversion efficiency from the two-oscillator element is identically 0 (black curve in Fig. S6(a)).

By moving the two resonances apart (say ~4.5 μm and ~7 μm) as in Fig. S6(b), the two contributions to the polarization conversion no longer destructively interfere (see S6(e) for the phase), and therefore there is substantial conversion intensity (black curve in S6(b)). Note that the movement of one resonance with respect to the other can be achieved in Y-shaped antennas by varying $L_T$ (Fig. S5). A similar effect is seen by keeping the two resonances roughly at the same wavelength, but changing their linewidths significantly, as in Fig. S6(c), such that the contributions are π out of phase at very long and very short wavelengths, and also exactly on resonance, creating the sharp dip in polarization conversion.

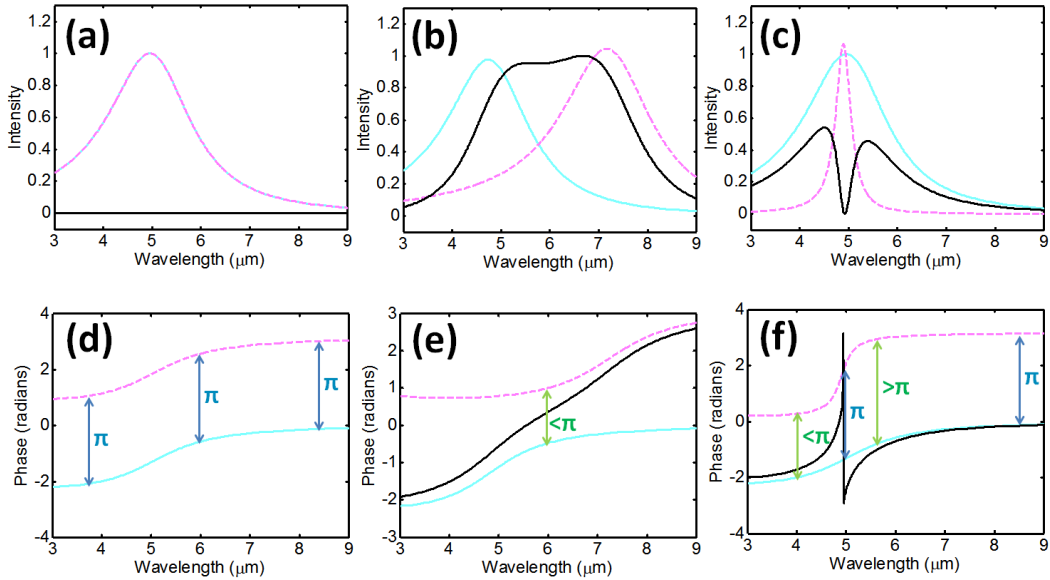

**Fig. S6** Polarization conversion due to the x- or y-oriented mode only (blue and pink curves, respectively, and due to both (black curve) (a). The two modes are exactly overlapping, generating no polarization conversion due to destructive interference. (b). The two modes are similar in linewidth, but have differing resonant frequencies, so some polarization conversion is seen. (c). The two modes are roughly overlapping in resonant frequency, but have very different linewidths, leading to some polarization conversion but a 0 exactly on resonance where destructive interference occurs.

We further explore the relationship between the two modes and the total polarization conversion efficiency by keeping the two modes at the same resonant wavelength and with the same linewidth, only changing their relative amplitudes. As can be seen in Fig. S6(a), when the two modes are identical in every way, the polarization conversion is 0 at all wavelengths (black curve) due to perfect destructive interference. By dropping the amplitude of one of the modes (pink) in Fig. S7(a), we obtain some polarization converting signal. This signal increases further and further with decreasing amplitude of the pink mode (Fig. S7(b)) until finally reaching the exact value of the contribution from the blue mode (Fig. S7(c)), since there is no longer any contribution from the pink mode. The phase response of both oscillators is plotted in Fig. S7(d), and is the same for (a-c). As expected, despite the $\pi$ phase difference between the two curves, complete destructive interference only happens when the two amplitudes are identical (as in Fig. S6(a)).

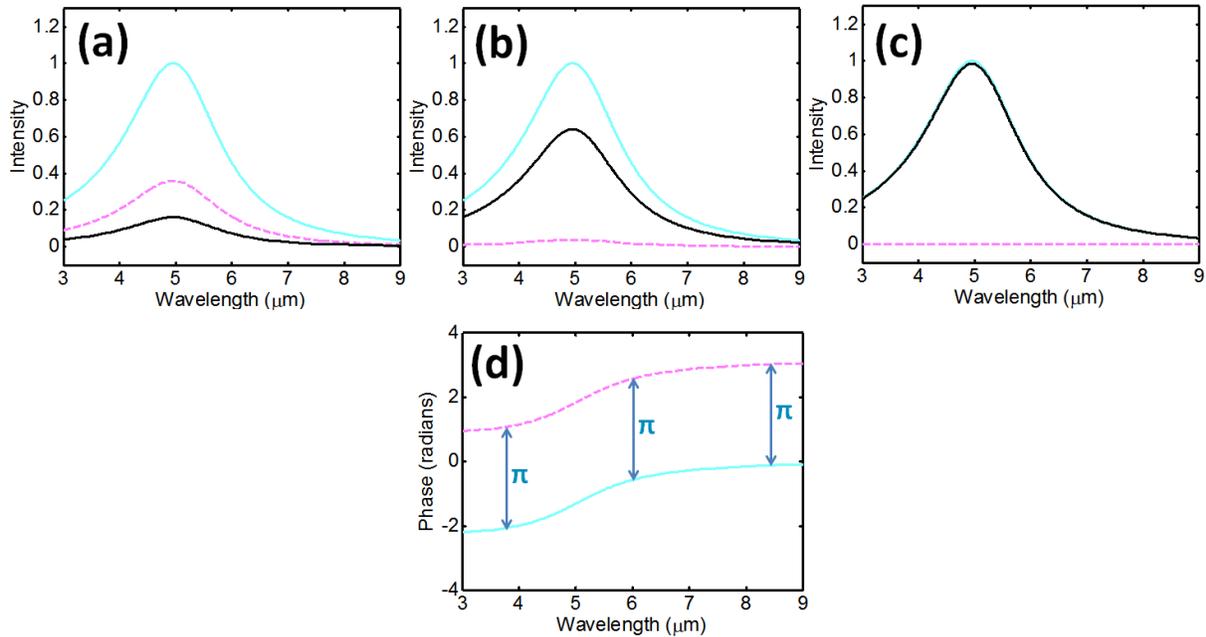

**Fig. S7.** (a) - (c). Polarization conversion due to the x- or y-oriented mode only (blue and pink curves, respectively, and due to both (black curve). The strength of the y-oriented (pink) mode is decreased between panels (a), (b), and (c), increasing the value of the overall polarization conversion. (d). Relative phase of the contributions to the cross-polarization of the two modes.

**Effect of a finite substrate**

The majority of the simulations shown in this report (including all of Fig. 2(a, b, c), Fig. 3(f, g, h), and Figs. S2, S3, and S4) involve metallic antennas placed on top of a silicon substrate, which takes up an infinite half-space, with the other half-space being air. In all of our experiments, however, the substrate is not infinite; instead, it is a 280μm-thick double-side polished silicon wafer. The finite extent of the substrate in the experiment significantly changes the spectrum of our metallic nanostructures. Because light can be reflected from both interfaces of the silicon wafer, a Fabry-Perot-type cavity is formed within the wafer, and the light which is trapped within this cavity affects the scattered fields from the antennas.

It is possible to account for this effect in FDTD simulations by simply introducing the finite substrate into the geometry. However, due to the differences in scale between the antennas (tens to hundreds of nanometers) and the substrate thickness (hundreds of microns), these simulations are exceptionally time- and resource-consuming. As a result, it is impractical to replace all of the simulations in Figs 2, 3, etc with ones taking into account the finite substrate. Instead, we performed one such simulation to correct the simulated spectrum of Fig. 2(e) and demonstrate how the substrate generally alters the spectrum of our antennas.

The resulting polarization-conversion spectrum is re-plotted in Fig. S8(a). The main effect of the substrate is the broadening of the antenna resonances and the washing out of the dip in the polarization-conversion spectrum. This effect is beneficial because it increases the effective bandwidth of our antennas, and can be explained as follows.

A substrate with finite thickness behaves as a Fabry-Perot resonator, which enhances the scattering efficiency of the antennas by feeding back some of the energy which would have otherwise been scattered away. In proximity of the resonance frequency of antennas, the scattering by the antennas is maximized, and since it is a loss channel for the Fabry-Perot resonator, the intensity build-up within the substrate is minimized. Therefore, the scattering enhancement provided by the Fabry-Perot modes is also minimized. At frequencies away from the antenna resonances the scattering cross-section of the antennas decreases, allowing more energy to be stored in the substrate. Consequently, this provides a larger enhancement to scattering from the antennas.

This effect serves to flatten out the scattering spectrum of our antennas, washing out features such as the dip around 4.5μm expected in presence of an infinite substrate (Fig. S8(a)). In the simulated spectrum that takes into account the effect of finite substrate thickness (Fig S8(a) black curve), the Fabry-Perot fringes have been low-pass filtered to reproduce the effect of the finite numerical aperture (0.4) of the objective in the experimental setup, along with the finite spectral resolution of the measurement. Due to the large order of the cavity (the thickness is much larger than the wavelength), small numerical apertures are sufficient to filter out most of the Fabry-Perot fringes. However, low-pass filtering does not eliminate the overall scattering enhancement provided by the Fabry-Perot modes. As shown in Fig. 2, this calculated spectrum matches well with the experimental data. For completeness, we show in Fig. S8(b) the expected spectrum for zero numerical aperture, where the Fabry-Perot fringes are shown, as well as the spectrum normalized to the average intensity (Figure S8(c)), which shows that the energy density inside slab (represented by the relative amplitude of the Fabry-Perot fringes) is minimized in the vicinity of the antenna resonances.

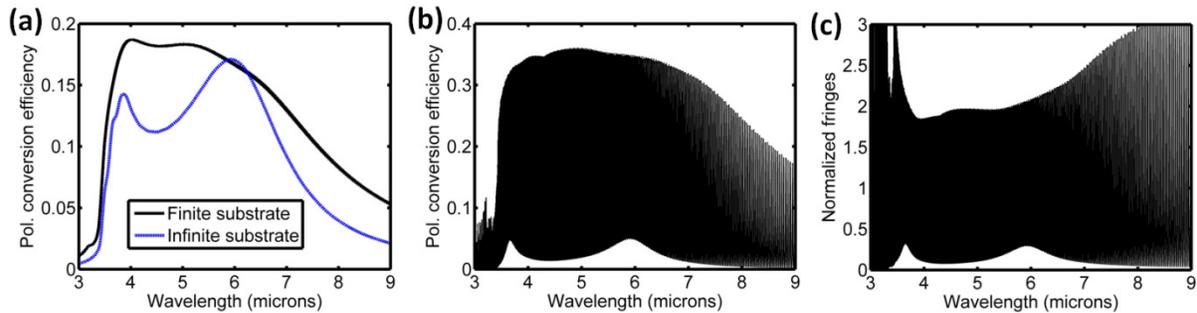

**Fig. S8** (a) Polarization conversion spectrum from a V-antenna array (reproduced from Fig. S2(e)) as calculated by FDTD with a 280μm-thick substrate (black) and an infinite substrate (blue). The finite substrate curve was smoothed with a low-pass filter. (b) Raw polarization conversion spectrum from the simulation without the filter. (c) The curve in (b) divided by its low-pass-filtered version which shows the relative amplitude of the fringes.

**V-antenna amplitude and phase plots at a constant frequency**

In the main text, it was stated that there were two complementary approaches to studying resonant behavior: fixing the resonant frequency and sweeping over the operating frequency, and the converse. The latter approach is difficult to take experimentally or with full-wave simulations because generating plots such as those in Fig. 2(a, b) involves the fabrication, measurement, and simulation of a different geometry for every value plotted on the horizontal axis. However, there exist techniques for calculating the amplitude and phase response by much more efficient means than full-wave simulations. In an upcoming work, we implemented the method of moments (MoM) for the efficient calculation of the resonant behavior of V-shaped antennas [S5]. In Fig. S8, we show preliminary calculations using the MoM of the amplitude and phase response of the scattered far-field in the normal direction to the interface, sweeping over geometrical parameters h and Δ (a, b) and over the wavelength and Δ (c, d).

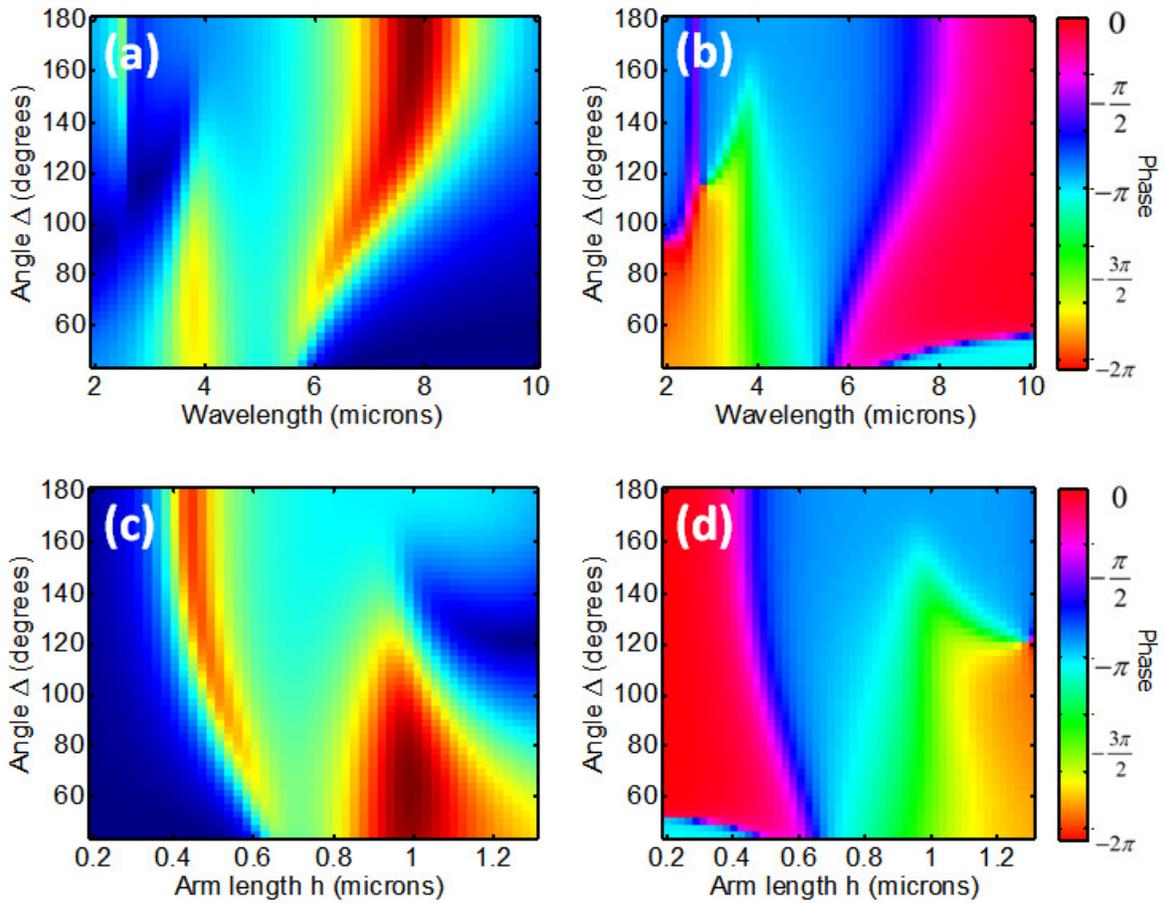

**Fig. S9** Approximate response of V-shaped antennas using the Method of Moments (MoM). (a, b) Amplitude and phase maps giving the phase and amplitude response of V-antennas in cross-polarization at a fixed arm length h ~ 600 nm, sweeping over the opening angle Δ and the incident wavelength. (b,c) Amplitude and phase maps at a constant wavelength = 6 μm, sweeping over the arm length h and the opening angle Δ.

These plots show that in general, the two approaches of analyzing V-antenna resonances are equivalent. Note that the maps in Fig. S9(a, b) corresponds roughly to those in Fig. S3(b, c), which were generated by FDTD. The maps are qualitatively the same, with the differences being due to the phonon resonance in the native oxide at 8-9 μm and near-field coupling to neighboring antennas, which were not taken into account in the MoM calculation. Additional, the MoM calculation included a thin-wire approximation (width << length), which does not strictly apply to our antennas, especially for shorter antenna lengths.